PACT Technical Report
No. 4                                                                                          Date:     29 October 2020

Subject: Realizing the Promise of Automated Exposure Notification (AEN) Technology to Control the Spread of COVID-19: Recommendations for Smartphone App Deployment, Use, and Iterative Assessment

Authors: Jesslyn Alekseyev, MIT LL; Erica Dixon, UPenn; Vilhelm L Andersen Woltz, MIT; Danny Weitzner, MIT

# Table of Contents

# 1 INTRODUCTION

During an infectious disease pandemic, it is crucial to actively manage disease spread to avoid overrunning the healthcare system in the time it takes to end the pandemic. One effective method to stem spread is to identify both infected individuals and those who came into contact with them, then encourage all identified to isolate and quarantine from other members of the population. Contact tracing is an important piece of the public health strategy for managing disease outbreaks but it is a largely manual and time-consuming process that is strained by COVID-19's high levels of asymptomatic spread, long disease incubation period, and high interconnection between individuals.

By using modern cryptographic techniques, privacy-preserving Automated Exposure Notification (AEN) technologies offer the promise of mitigating disease spread by automatically recording contacts between people over the incubation period while maintaining individual data privacy. AEN has the potential to identify cases traditional contact tracing cannot account for, such as contacts though public transportation, social events such as concerts, or indoor dining at restaurants. If AEN apps are used throughout populations with in-person contact, they could serve to improve the efficacy of contact tracing methods, helping to keep infection rates low and preventing additional waves of infection.

Today, public health departments in States and other countries around the world are deploying AEN systems at a rapid pace. Though many organizations conducted research prior to deploying apps, experience around the world shows that contact-tracing apps are installed and used at relatively low levels. Existing research suggests that barriers to user adoption include: concerns about privacy and security even with apps leveraging privacy-preserving technologies such as the Google Apple Exposure Notification (GAEN) system; low levels of trust in medical systems, government, and/or tech companies; as well as doubts about the efficacy of the technology. Additional factors include the potential that some user groups are not tech savvy, or do not feel the need to take extra precautions during the pandemic.

States considering the deployment of AEN apps need to set expectations appropriately. While the promise of AEN technologies is there, there is still work to be done to realize its full potential. There is a learning curve with every new technology while industry and researchers work to understand user information requirements as well as to combat fear and anxiety among the general population. As long as States work to ensure data protections and maintain trust, there is room to iteratively test and improve existing apps. For States still considering the deployment of apps, there is significant opportunity to learn from existing deployments. States are encouraged to leverage human factors and user experience (UX) research methodologies to understand user experience with technology and information, and investigate those aspects with the goal of improving technology adoption and use. These methodologies give us tools to understand what barriers to adoption exist and how we might overcome them.

This whitepaper is intended to provide usable information for States who are considering the deployment of an AEN system, as well as to guide ongoing improvements for States that have already deployed. We outline the human factors considerations related to employing AEN systems with the ultimate goal of controlling the spread of COVID-19, including the GAEN consortium Exposure Notifications (EN) Express tool. We will also provide a practical design and implementation guide for States and others designing and deploying AEN systems, as well as a set of recommendations for assessing deployment of contact tracing apps and targeting areas of concern to improve efficacy of use during and after initial deployment. As a case study, we consider the commercial app deployed by the state of Pennsylvania (PA) and the ongoing efforts to drive user adoption there.

# 2  AEN SYSTEMS AND CHALLENGES

States who have deployed or who are considering deployment of an AEN app have a number of considerations. First, they need to decide on the technology and method of deployment. Here, we will discuss the trade-offs of different "flavors" of AEN system deployment. We will then discuss user experience with AEN technology as a whole in order to provide a comprehensive overview of the problem space.

## 2.1  APP SELECTION AND DEPLOYMENT

There are multiple deployment methods available for States interested in AEN apps; **Table 1** shows options available to States with a few of the considerations for comparison. Note that we only discuss relative challenges; we do not recommend a specific course of action. To date there is no clear "winner" or magic formula, though there is some indication that better integration between legacy and new contact tracing systems as well as the flexibility to integrate lessons from ongoing research into messaging in and around the app may improve user experience and app use.

*Table 1:* General AEN App Deployment Methods with Selected Comparison Points

| Method | Development Cost | Technology Choice | Customization | Standardization | Data Access | Message Control |
|---|---|---|---|---|---|---|
| Deploying a commercial app without modification | Moderate - Low | Moderate | Low | Varies | Moderate-Low | Moderate |
| Deploying a commercial app with modification | Moderate - High | Moderate | Moderate | Varies | Moderate | High |
| Designing and developing app | High | High | High | Varies | Moderate | High |
| Using EN Express without an app | Low | Low | Low | High | Moderate - Low | Low |

Many app developers have chosen to leverage the Google Apple Exposure Notification (GAEN) system announced by both Apple and Google in April, 2020 (Google, 2020), with 19 States as of this writing choosing to deploy apps that leverage GAEN technology (Hall, 2020). The Google/Apple agreement includes an API that allows apps from public health organizations to communicate between phones running Android or iOS, and offers a Bluetooth-based, privacy-protecting automated exposure notification capability. Developers that leverage GAEN technology do not have to develop or vet contact tracing technologies themselves, essentially leveraging Google and Apple development work to speed up app development while ensuring a high level of standardization with other apps leveraging the technology. All deployment options in the table above have the potential to leverage GAEN.

**Deploying a commercial app without modification.** States can choose to leverage an existing app by either recommending an existing app from Google or Apple's app stores to their state population, or by contracting with an app developer without modifying the technology or interface presented to the user. States can also choose to minimize the interactions between the app infrastructure and existing contact tracing and public health infrastructure to minimize cost further. However, this cost reduction comes at the expense of flexibility, including the ability to adapt to state population needs, or to lessons from States that may find ways to improve the impact of their deployments. States are also tied to app developer technology choice. While this allows States to choose an app with preferred technology performance, the underlying technology may or may not be standardized across apps and other States. States may also have limited access to data around app deployment, which also limits opportunities for evaluation and improvement. States can always choose how to advertise the app, though messaging coordination between manual contact tracing infrastructure, public health, marketing, and the app may be difficult. The CDC provides guidance for States surveying existing apps: https://www.cdc.gov/coronavirus/2019-ncov/downloads/php/guidelines-digital-tools-contact-tracing.pdf

**Deploying a commercial app with modifications.** Many States have selected this option for good reason. While it may be more expensive than deploying an existing app as-is, it allows States the flexibility to customize in-app messaging and branding, as well as coordinate with existing contact tracing and public health infrastructure and messaging. States can also choose to implement some data collecting screens or questions in the app to improve their data collection, though app users would need to opt-in (see the discussion of **In-App Trustworthiness Principles** below).

**Designing and developing custom apps.** Some States have opted to contract with an app developer and develop a custom app. Though this affords States a high amount of customizability and flexibility to coordinate with existing state infrastructure and messaging, and for States to leverage their preferred tracing technology, this approach can also be challenging. States that do not have the time, resources, or funds to fully investigate user requirements or to properly vet available technologies and vendors are accepting a high amount of risk. States are still limited by Google and Apple data collection allowances as long as the app is in an app store, so data collection ability is similar to deploying a commercial app with modifications.

**Using EN Express.** The GAEN consortium has a free notification capability that States can opt into that leverages GAEN. While this Exposure Notifications (EN) Express system provides automated notifications, it does not provide other user-focused features such as state-wide metrics, links to state resources, or symptom trackers that provide additional value to users. As with deploying a commercial app without modifications, States can choose how to advertise the app, though messaging coordination between manual contact tracing infrastructure, public health, marketing, and the app may be difficult.

Once States deploy an app, they should anticipate low download and usage rates without significant advertising and/or messaging campaigns aimed at getting the word out to users across the state. In order to fully realize the potential of AEN technology, States should also anticipate needing to assess their deployments and adjusting messaging and/or app functionality over time.

## 2.2 TECHNOLOGY AND USER INTERACTIONS

States who have deployed apps are currently working together and with universities to gain insight into improving download and usage. Anecdotal evidence points to low awareness or understanding of how contact tracing works or provides benefit throughout the general population; adding a technology component is expected to compound that. Leveraging the ongoing work to improve AEN app usage or contributing to it requires an understanding of how the technology works, as well as how users experience technology and interfaces.

Any time a new technology is introduced, it has to contend with the existing socio-technical system of people, social structures, technology, and tasks (Bostrom and Heinen, 1977). **Figure 1** describes the environment surrounding app usage. At left (User Decision Making and Environment), users bring their past experience, bias, goals, and understanding of tasks to any interaction with an interface. From right (Technology Capabilities and Structure), the interface also acts as a translation layer for the underlying technology and implementation/code (Alekseyev, 2019).

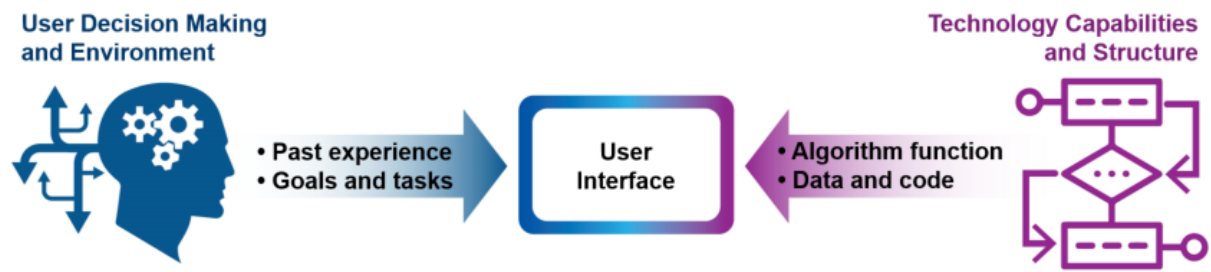

*Figure 1: Interface usage environmental considerations. Users bring their past experiences to the interface, and use the interface to understand the underlying technology.*

**User Decision Making and Environment.** When considering AEN app use, users bring their understanding of both the term "contact tracing" and the act of collecting data for use in contact tracing efforts, as well as any bias towards the term, COVID-19, marketing, and the organizations associated with the app to the interaction. This understanding may not be accurate, and may not reflect the intent of any organization involved.

Primary barriers to user adoption are likely to fall into three main categories: mistrust, uncertainty or confusion, and disinterest. Mistrust toward these apps may stem from a variety of places, including suspicion toward the government (local, state, or federal) and its goals in using the app, concern about large tech companies and their involvement (Binns et al, 2017), or hesitations about privacy and security of user data, including location and medical information. Many States have eliminated language focused on contact tracing or tracking from app descriptions, both in order to alleviate user concern about personally identifiable or location based data being collected, and because the apps do not work as manual contact tracing does (i.e. traditionally, known contacts to positive case are contacted by tracers; here, exposed cases are only alerted through the app, and not known to public health authorities).

 Instead, app descriptions focus on exposure notifications with Google phones containing a setting for "COVID-19 Exposure Notifications", while Apple iOS refers to their setting as "COVID-19 Exposure Logging". While these moves are expected to help promote user trust and encourage engagement, there are still many unknowns about the short and long-term effects these measures will have on user adoption, and whether additional measures should be taken. There is the additional complication arising from the deployment of Google and Apple settings to their phones; a cursory review of social media sites after their deployment uncovered a number of posts warning people about the new "tracking" that Google and/or Apple snuck into the phones while people were unaware.

Uncertainty or confusion about these apps is likely to be a barrier in any state, and has been prevalent in States where apps have been released. Social media users have concerns about location data being collected, predominantly focused around concern that the government is tracking users through the app, but also concerns about people deliberately reporting false positives to cause panic. Finally, disinterest in downloading the app may be driven by doubts about efficacy of the app (i.e. that not enough people will download the app, making it functionally useless), disbelief in COVID, or simply a lack of awareness about the app.

Barriers to user adoption can stem from the state side, including a general lack of understanding of particular barriers in their state in general, among the population they are most trying to reach in their state in specific, a lack of identifying strategies prior to launch, or not continuing to evaluate barriers post-launch as they change. For example, the biggest initial barrier may simply be lack of knowledge that the app exists, but ongoing barriers may include negative app experiences, such as intrusive notifications or bugs that cause people to delete the app. Additionally, there may be a number of assumptions about who will and won't participate in these groups that may be informing initial roll-out of marketing and outreach, that States should work to validate or adjust.

Though there are many barriers stemming from user biases towards the technology and organizations involved in the technology, there are many opportunities to influence app experience and use by working to understand existing user attitudes, and shaping the information presented to the user through messaging around app deployment (see **AEN App Assessment Guidelines**). There is also significant opportunity to research these issues and develop mitigations.

**Technology Capabilities and Structure.** In order for AEN apps to affect virus transmission rates and assist in controlling the spread, the technology does need to function as intended, but state populations also need to be willing to download the app and engage with it appropriately. Being mindful of how the system of user-and-technology can break down is important to determining interventions that can improve effectiveness. As the GAEN system is quickly becoming the technology of choice with 19 States indicating use of or intent to use as of this writing (Hall, 2020), we will focus on describing the general workings of the GAEN system, i.e., AEN technology leveraging phone Bluetooth capabilities.

As shown in **Figure 2**, a person positive for COVID-19 (Index Case in red) needs to have been carrying an AEN app, which requires that they have a smartphone, locate the app, download it, open it, have Bluetooth enabled on their phone, left their phone on, and left the app running. The app produces a number of anonymized Bluetooth "chirps", and records chirps that it comes across, which requires that it come within sensing distance of another phone that also has the app enabled. The app does not know if the Bluetooth signal is blocked by material, either by being in a pocket or a purse, or if it is near another phone as opposed to a person carrying a phone. It also doesn't distinguish between household members and strangers, or whether anyone was wearing a mask – it simply senses the presence of other Bluetooth chirps where it can, and stores those locally.

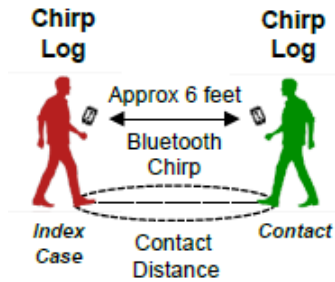

*Figure 2: Typical function of Automated Contact Tracing (AEN) apps leveraging Bluetooth, where phone Bluetooth "chirps" from an index case (initial positive case) is sensed by a contact who has been "too close for too long," typically defined as within 6 feet for 15 minutes or more.*

In order for the contact-tracing part of the app to work, someone with the app needs to test positive for COVID-19 (Index Case) and be given a code to type into the app. They then need to type the code into the app, the app needs to successfully send the chirps it has recorded over the last 14 days to a central server where they are compared automatically. Any phone where a "too close for too long" contact has been identified (typically defined as within 6ft for 15 minutes or more) will receive a notification. In order for this notification to affect virus transmission, the new phone (Contact) would need to receive the exposure notification, and the user would need to follow the instructions provided by their state which may include quarantining as well as potentially getting tested. If they receive a positive test (becoming a new Index Case), they must receive and upload a code to continue the process.

Technology selection can impact user experience, compliance, and engagement. For instance, the EN Express system minimizes the need for people to locate and download apps as it can be deployed automatically to some phone users when they upgrade their phone operating system, though other technology limitations and considerations still apply.

There are opportunities here to affect user experience and effective app deployment. Communications in and around the app can be employed to ensure all app users understand how to use the app effectively, and to ensure that users understand how to leverage the app technology appropriately. For instance, users may not be aware of Bluetooth limitations, and may mistake lack of notifications due to a phone being buried in a backpack or lack of others having the app for overall lack of COVID-19 exposure. Also, States should consider focusing efforts on populations that are typically in contact with people outside of their direct household (see **Success Metrics Considerations**). There are also some indications that users would find more value in preventative rather than reactive measures; investigating phone features that provide needed information for users to keep themselves safe may improve opinion and use. See **AEN App Assessment Guidelines** for specific assessment recommendations.

**The Interface.** The interface serves as a translation layer between the underlying technology and the user, and can serve to help or hinder user adoption. There are opportunities to influence app experience, use, function, and understanding by shaping the information presented to the user through each screen they view and each message they receive. States should also consider other messaging available through the phone; separate apps likely have to contend with messages pushed to phones by Google or Apple outside of the app itself. Though the difference between the app and the notifications may be clear to States as well as Apple and Google, it may not be so clear to phone users. States that consider other environmental messaging in their app evaluation and messaging efforts may realize more successful app use. Some States, including PA, have additionally seen indications that additional informational features such as the ability to see where increased spread is occurring would improve user experience with the app.

## 2.3  IN-APP TRUSTWORTHINESS PRINCIPLES

Here, we provide an overview of several design principles and specific recommendations to work towards increasing user trust in and engagement with automatic contact tracing mobile apps. Existing applications should be evaluated using the principles described, in order to identify means to improve app communications and user experience. See **AEN App Assessment Guidelines** for specific assessment conduct recommendations.

Users' willingness to provide a mobile app with their personal information depends on a variety of factors that influence users' perceptions of trustworthiness. In addition to the app-external aspects such as the relationship of the user to the entity operating the app discussed above, there as aspects of the app's operation, such as what information is requested, how it will be used, who will use it, visual design, content, language, and tone in communication.

When designing any app, user trust and engagement can be increased by considering various factors, including:
- being fully transparent regarding information collection and use
- giving the user full control over their information
- minimizing requests for user action
- having a simple visual layout
- minimizing information displayed in notifications

We will now turn to specific design and operation recommendations for individual discussion.

**Maximize transparency and user control.** Care should be taken to explain every need and use of personal information, and control should be given to the user to decide to share each individual piece of information (IAPP 2012; FTC 2013; Liccardi, Weitzner, Pato 2013). The app should be designed as to avoid any unnecessary conditional requirements of information to achieve a goal. For example, whenever avoidable, users should not be required to fill out forms that require specific personal information to proceed to the next step of some process.

This principle holds throughout app use; in particular, after initial steps have been taken it should remain easy to make any changes to user-information permissions that a user wants to make. The control of each user over their personal information should be maximized to build trust between the user and the operator of an automatic contact tracing app. With AEN apps, this principle means that users should retain control over all additions of personal information to the app, including name, address, phone number, and information on when a user tests positive.

**Minimize requests for user action.** In a similar vein of user control, requests for user action should be kept to a minimum (FTC 2016). As much as possible, any action within the app should be left up to the user to decide whether to take that action or not. This gives the user control over their use of the app without recurring prompts for action or information if the information is not required for basic functionality.

The app should never be able to prompt the user to upload a positive test result. The app should additionally minimize all requests for user information; each prompt should be critically analyzed to determine if it is necessary or not.

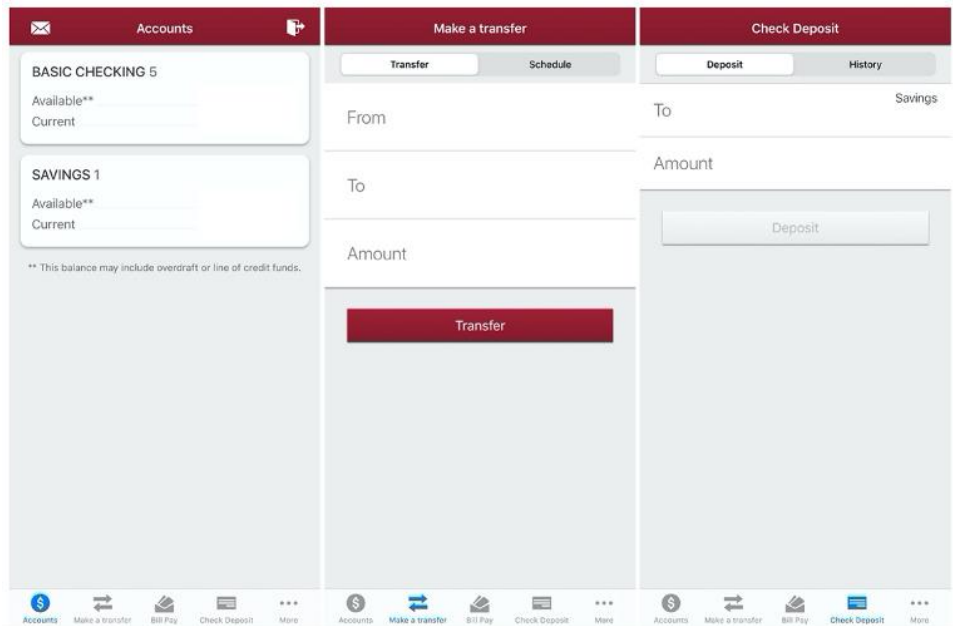

*Figure 3: A simple banking app layout. These mobile banking app screenshots, taken from the MIT Federal Credit Union mobile app, are a great example of a simple layout that allows the user to easily control each action they may or may not take within the app.*

**Simplify app layout.** The app layout should be kept simple and clear to increase engagement and ease of use (FTC, 2016). For example, banking apps, which are widely entrusted with very important and private information, commonly have a very simple and minimal layout – they show no more than what is necessary to provide specific information or guide the user as they accomplish specific actions, such as a transfer between accounts or a check deposit (*Figure 3*). Simplification of the layout helps the user understand the app in its entirety and allows the user to control their interactions with the app more easily and effectively (Liccardi et al, 2014).

**Carefully send and write notifications.** Notifications should be used sparingly, and only encourage the user to check the app for any specifics for why the user is being notified. For example, returning to a finance or banking app example, typical notifications are sent only to inform the user that a statement is available for them to read in the app. A notification of this form implicitly assures the user that the entity is only willing to share information once the user is logged in, which safeguards inadvertent leaks of any personal information that might be in that notification (FTC, 2016).

Specifically, if a user has had a positive contact, the app should not directly display that information in a notification. Instead, the app could notify the user to check the app for details regarding a notification, and only once the user is logged into the app should it reveal that the user was in contact and provide information regarding available support systems.

In the next section, we'll discuss how to evaluate user experience with the principles described, and determine means to bring apps further into alignment with these principles.

## 3  HUMAN-FOCUSED ASSESSMENT METHODOLOGIES

A number of methods are appropriate to apply to evaluation and iterative improvement of AEN apps, both to analyze usability issues and develop design solutions to address those issues.

### 3.1  METHODS OVERVIEW

A wide range of methods are in use by human factors and usability researchers today, resulting in any number of potential research plans or means of conduct. Conducting comprehensive studies can require a lengthy design review cycle. However, given the urgency of the current public health crisis, methods need to be adopted and rapidly deployed to keep pace.

**Table 2** shows several methods, the types of questions a method is best at answering, what data it is best at gathering, and what level of effort and time commitment each method requires.

*Table 2: Assessment Methods Comparison*

| Method | Identifies | Qualitative | Quantitative | Behavioral | Attitudinal | Effort | Time |
|---|---|---|---|---|---|---|---|
| Surveys | How many? | some | X | | High | Low | Low |
| Focus Groups | Opinion (concepts) | X | | | High | Moderate | Moderate |
| Interviews | Why? | X | | | High | Moderate | Moderate |
| Design Workshop | Design hypothesis | X | | some | High | Moderate | Moderate |
| Heuristic Review | General usability | X | | some | some | Low | Low |
| Usability Testing | Usability, Why? (once) | X | X | High | some | Moderate | Moderate |
| Diary Studies | Usability, Why? (over time) | X | some | High | some | Moderate-High | Moderate-High |
| A/B Testing | Design, usability | X | X | some | High | Moderate | Moderate |
| Analytics | What? How many? | | X | High | | Low | Low |

We discuss each method in more detail below:

- **Surveys\*** rely on responses to a pre-set series of questions. While surveys can be used as a qualitative measure, they are well-suited to gathering quantitative data, allowing for rapid deployment to a large number of people. This method relies on what people say, which may not translate to behavior. For instance, with AEN, many users indicate that they care about data security when asked, but their actual behavior changes depending on the data they are being asked about, as well as how the information presented to them (Shih, Liccardi, Weitzner 2015). States should consider asking only about information related to actionable changes that could be made (e.g. design of app interface is changeable, while switching from Bluetooth to location based tracking is not), and supplement surveys with behavioral data to understand the *why* behind survey answers.

- **Focus groups\*** are discussions with a group of people run by a moderator intended to tease out different reactions and opinions about products and ideas. This method generates qualitative data and can generate important information about group behavior and dynamics, though it can also mask concerns and opinions of individuals. This method also relies on what people say, and should be supplemented with behavioral and quantitative measures.

- **Interviews** are typically one-on-one discussions with individuals. While this method is also focused on user attitudes and generates qualitative data, interview questions can be focused to probe *why* people feel a certain way, which is difficult to do in a survey. It also helps researchers probe people's opinions without being influenced by group dynamics, and can help researchers explore user attitudes deeply to inform quantitative data gathering methods.

- **Design workshops** are group discussions focused on framing problems and generating design solutions with stakeholders. The method generates qualitative data focused on understanding the design space. Workshops can help people frame their ideas and discuss concerns in a way that has direct impact on the design solution, minimizing translation issues between questions posed by moderators or interviewers and design solutions intended to address concerns.

- **Heuristic reviews** are evaluations of an interface or product conducted by an individual or a small number of people taking on the role and perspective of an intended user. Reviewers can use generally accepted usability heuristics (Nielsen, 1994) to structure feedback. These reviews only require a reviewer and an interface to produce qualitative information focused on interface improvements. This method is a good way to discover general usability issues quickly, but can miss specific experience requirements of different populations. This method should be combined with other methods to limit the risk of focusing app adjustments on the preferences of the reviewer

- **Usability testing** involves recruiting intended users to walk through realistic tasks with an intended interface at one point in time, resulting in information about the usability of the intended workflow as well as information presentation.. Testing can be conducted with any level of design fidelity from sketches through deployed software. Remote, unmoderated tests can gather quantitative data that includes some information about behavior and preferences. Moderated, one-on-one sessions allow moderators to explore user experience and behavior at a deeper level whether that is conducted through web conferencing software or in person.

- **Diary studies\*** involve recruiting intended users to record their impressions of an interface over time. As this requires users to stay engaged with the data collection request for a longer period of time, the studies can record behavior in a more natural setting, and allow researchers to understand changes in behavior or experience over time.

- **A/B testing** involves recruiting intended users to compare different design solutions, and is used to gather user feedback on multiple design hypotheses in one testing session. As with usability testing, this method also can be deployed in a number of different ways, with similar strengths and limitations. It also results in qualitative or quantitative data, depending on the method of deployment.

- ***Analytics** involves gathering statistics that are collected either in-app, through the app store provider, or collected outside of the app to evaluate usage. This method results in quantitative data that should be considered together with qualitative measures.

*Both marketing and human factors researchers use this method, and there can be some overlap in the type of questions asked and information gathered. We recommend coordinating human factors and marketing research as early as possible to ensure teams are learning from each other and leveraging each other's research.*

## 3.2    SELECTION AND APPLICATION

Deployment teams are encouraged to use a variety of methods to investigate different aspects of user experience, as well as to vary facilitators, observers, user groups, geographic areas, or other aspects that may skew research results (Wilson, 2006). Ideally, researchers should choose methods that cover different aspects listed above, such as selecting both attitudinal-focused and behavioral-focused methods, and taking steps to reduce the bias introduced by one researcher's perspective as well as one method's strengths and weaknesses.

Lean UX methodology provides a guide for how to conduct research in a rapidly changing environment, such as in startup companies where time-to-market may mean the difference between success and failure. The goal of Lean UX methodology is to identify the minimum number of capabilities needed by users, and to use lightweight research methods to iterate and push viable capabilities out to users rapidly (Gothelf, 2013). Practitioners rely on consistent user feedback throughout development to drive project development and reduce risk. States and teams that adopt an iterative application of research with a focus on getting capabilities to users quickly may realize a positive effect quickly, with many opportunities to iteratively assess and improve experience.

However, speed-to-market needs to be weighed against quality and user perception. Deployment teams need to take care that they thoughtfully craft and apply research plans to further project goals; research without proper considerations in place and proper interpretation of results can misguide projects (Cockton & Woolrych, 2002; Greenberg & Buxton, 2008). Projects that are rushed to market without proper vetting have the potential to drive users away and make it harder to bring them back, particularly where potentially sensitive data is being collected. Though missteps in deployment can be corrected if an iterative research and improvement plan is in place, thoughtful integration of use research and proper interpretation can help prevent missteps from occurring.

## 3.3    SUCCESS METRICS CONSIDERATIONS

A major roadblock to understanding whether these apps are successful is a lack of consensus on what success means. A frequently referenced Oxford University model posits 60% of the population downloading the app as the critical number for stopping the pandemic, which caught the attention of many – some of whom used it as evidence that these apps would not be successful (University of Oxford, 2020). Such judgement, however, is premature. Stopping the spread of disease is the strongest possible benchmark to achieve, and there are concrete benefits that can be achieved with lower adoption rates. The Oxford team stated that even with a lower number of users there would still be a reduction in cases and deaths; this claim is supported by evidence from Washington state indicating that with 15% of the population downloading the app, case counts and deaths could reduce by 8% and 6%, respectively (Abueg et al, 2020 [preprint]), as well as simulation results that show that even 20% of the population using the app has an beneficial impact on the spread of the virus due to the relative speed with which AEN methods trace contacts.

In addition, the "60% of the population" level of adoption standard could be applied to more granular views of what constitutes the entire population. For example, if 60% of a given state population downloaded the app, but that population is composed of people who are working from home, sheltering in place, or social distancing, the app does little to control the spread of COVID-19. Instead, if even 20% of a population of essential workers or of college students living on campus downloaded the app, that could have much higher impact despite the lower rate of downloads. For properly chosen groups, community saturation may be much more critical than overall population saturation. It may be that we need to reparametrize these models with the percentage of people who are in sustained (15+ minutes, <6ft apart) contact with other people who may be a source of COVID-19 exposure.

A second roadblock to quantifying app success is limitations in available data. Part of the privacy and security of the GAEN system prohibits the use of tools such as Google Firebase, which generates reports of the age, gender, location, and interests of app users. Because of this, app store metrics offer no information about the demographics of users of the app. States should be aware of this while deciding on success metrics; if demographics are needed, they will have to be acquired through optional questions within the app, or through a linked survey.

In the next section, we discuss how to bring these considerations together to form an assessment plan.

## 4 AEN APP ASSESSMENT GUIDELINES

In this section, we use the above contextual considerations as well as work conducted with the Commonwealth of Pennsylvania (PA) to detail and discuss a recommended iterative assessment plan. See https://www.health.pa.gov/topics/disease/coronavirus/Pages/COVIDAlert.aspx for information about PA's deployed app.

### 4.1 SET MEASURABLE GOALS

Prior to the steps and related communication points above is a step 0 – determining which app deployment to use, and setting initial measurable goals to achieve. If measurable goals aren't set at the outset, they should be set at some point during app deployment to give States a chance to measure "how are we doing?" and "how can we improve?"

Now that multiple States have released apps, States considering an app are well-positioned to evaluate available information, including app store reviews, social media conversations, news stories, and the released apps themselves to evaluate what is and is not successful for other States.

**Prior to app deployment.** We recommend that States set goals for AEN use, and take time to understand what will work within their state budget and population. States planning to implement an app should start by understanding what can be addressed prior to launch, such as measuring app acceptability among the population through various measures, and what needs to be consistently addressed through ongoing evaluation. States should work to understand where it is possible to mitigate fears with messaging and design, as well as what is likely to do with factors that go beyond technical design. States should also coordinate with their public health entities to understand sources of spread in the state, as well as vulnerable populations, and consider targeting app communications towards those communities. Each state should have a solid idea of what they want to get out of app deployment and what they hope to affect by deployment; this understanding will drive efforts to assess and improve app deployment effectiveness.

In PA, there were a number of steps taken prior to app deployment. First, the state determined that they would use the developer for Ireland's app, NearForm, and adopt the app for use in the state. The app would be branded for PA use, but the general function, information, and underlying technology (GAEN) would remain the same. This would ensure that the app could be rolled out relatively quickly, and took advantage of extensive attitudinal/behavioral and technological testing that had been completed by the NearForm team.

The state additionally determined an initial pilot and roll-out schedule, where the app function would be tested with a smaller audience before rolling out state-wide. Participants for the opt-in testing during pilot included PA state employees in the Pennsylvania Emergency Management Agency (PEMA) and the Pennsylvania Department of Transportation (PennDOT), as well as a group of interested faculty and staff at the University of Pennsylvania (UPenn).

**Iteratively throughout deployment**. Setting initial metrics is just the first step. Assessments will help States assess the initial launch, though there may be many opportunities to adjust messaging or impact app use throughout deployment. States should consider assessing metric collection results as well as user experience and attitudes throughout the deployment of an app in order to adjust to changing conditions. This includes adjusting metrics, if the results from initial assessments are not meeting state needs.

**Recommended assessment methods and examples:**

- **Survey of state population.** Surveys are a low-cost instrument to take the pulse of sentiment across your state and to test hypotheses app developers or health officials have about public sentiment. When issued before roll-out, they can help inform marketing campaigns, as well as direct funds and efforts. After initial roll-out, they can be used to measure any change in sentiment, as well as the effect and reach of marketing campaigns to adjust or tailor efforts.

    What are people saying about COVID 19, contact tracing, AEN apps across the state? What are the vulnerable populations? What areas will have the most impact? The more specific you can make your targets and goals for AEN app deployment, the easier it will be to make design decisions and focus efforts; "10% uptake" is good, but "20% of the on-site college population" is better. Care should be taken to represent all demographic groups throughout the state, with particular focus toward groups experiencing COVID-19 health disparities and disparate rates of spread.

    Prior to app launch, PA conducted a marketing survey to gauge user attitudes towards COVID-19, as well as attitudes towards prospective advertising campaigns to focus messaging efforts. Researchers leveraged the results from the marketing survey to inform additional assessments as well as to drive initial marketing campaigns.

    During the one week pilot, PA users downloaded a beta version of the app, and were asked to respond to a survey about user experience at the end of 7 days, as well as testing the entry of test

codes and subsequent exposure notifications. Information from the pilot was used to make changes to the app, including adding an optional alert to prompt interested users to complete daily symptom tracking, and modifying entry language to be more accessible.

- **Heuristic reviews.** Many States and countries have deployed apps already, with varying rates of success. Examining existing app workflows and messaging from the perspective of an intended user can help States identify potential issues early, and can provide States with insight into what data the app is collecting and how to inform their metrics evaluation post-launch.

  Prior to app launch, researchers from UPenn, MIT, and MIT LL conducted a heuristic review of app workflow, in-app messaging, and information presentation. A number of requests were made to app developers to address expected misunderstanding of messaging, and potentially confusing interactions. The exercise also helped PA identify where the app was gathering data in context of use - for instance, when and where users were asked to opt in - to inform metrics gathering and assessments.

- **Workflow assessment**. Prior to app launch, researchers at UPenn, PA DOH, and elsewhere evaluated app technical performance and the overall workflow between the app and existing contact tracing and public health structures. This helped researchers to identify metrics that could be collected throughout the process of contact positive cases in order to develop an assessment plan to evaluate app effectiveness.

  The primary goal of any state utilizing exposure notification apps is likely to be total number of downloads; this is both an easily accessible metric via download counts and reflective of the idea that the more people who have the app, the more successful the app can be. However, PA also identified other goals and determined how to measure information needed to assess them:

  o Goal: Onboarding process is understandable and acceptable to users; Metric: Low number or percent of drops during onboarding

  o Goal: Demographics of people downloading app will match to population; Metric: Numbers from optional in-app demographics and optional in-app survey (both subject to sampling bias, but primary source of information)

  o Goal: Users provided with codes will enter and upload data; Metric: number of codes provided/number of codes entered

- *Optional: Focus groups, interviews.* Individual and group discussions can help round out information gathered by other means. If States have the resources to supplement surveys and reviews prior to app launch, talking to contact tracers, nurses, doctors, and potential app users individually, in groups, or both can provide insight to inform survey questions, as well as improvements to messaging in and around the app.

  While PA did not conduct focus groups or interviews prior to app launch, there are plans to integrate potentially both to gather data to drive iterative improvements in messaging and/or in-app components.

- *Optional: Analytics.* People chat online about things that are important to them. Reviews of existing apps can provide insight into what people are experiencing, as well as likes and dislikes, though States should take care to triangulate sentiment with surveys and interviews – each platform gives a snapshot of users of that platform who comment actively, and not necessarily the state as a whole.

Researchers at UPenn are conducting ongoing data mining analysis of both PA's and other States' app store reviews, in order to understand what drives positive and negative experiences with these apps. Researchers are also analyzing data from Twitter and Reddit to understand barriers to downloading, pervasive concerns or misinformation, and overall spread of relevant information.

## 4.2   FOCUS EFFORTS ON CONTROLLABLE COMMUNICATION POINTS

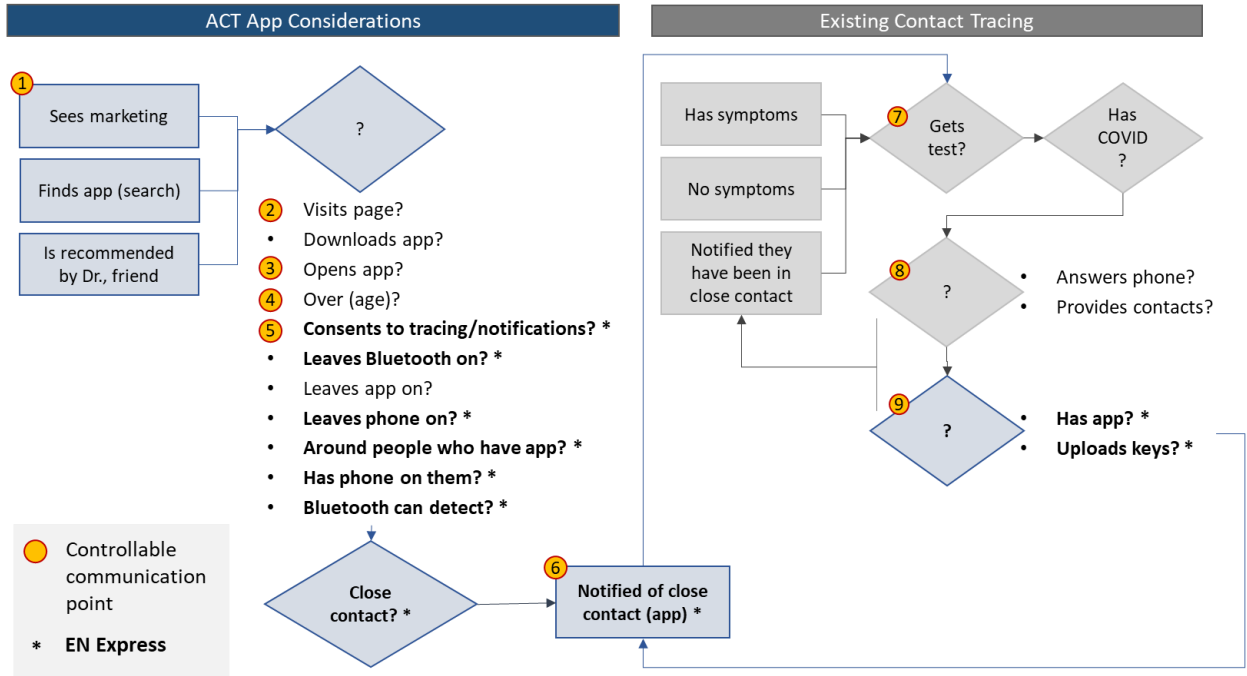

*Figure 3*: *User-focused contextual view of AEN app function and controllable communication points. Each step and question indicates a potential user drop-off point. There are a number of points where there is an opportunity to affect user experience and perception (numbered circles).*

**Figure 3** Shows the environment of AEN use from the user's perspective to describe interactions between users, AEN apps, and manual contact tracing structures. Each question in the figure indicates a potential user drop-off point, both with traditional contact tracing steps (gray), as well as AEN app use (blue). For each contact point (orange circles), there is an opportunity to affect user experience and perception. Details of in-app information and order of consent requests differ from app to app; bold items indicate what items also affect EN Express.

**(Controllable Communication Point 1) Marketing and outreach.** The first point of state communication with prospective users is through marketing and outreach. Here, it is important to consider that app-related messaging is competing with messaging prospective users are getting from friends, family, interactions with health care workers, social media, and elsewhere in a time where attentional resources may be stretched thin. Though States cannot control gossip or even Apple/Google in-app communications, they can take steps to foster positive engagement towards the app and messaging. Many organizations are avoiding the words "tracking", and instead are focusing on language around being awareness and notifications, with University of Arizona's app being branded "COVID Watch", and PA's app "COVID Alert" though as mentioned earlier, it is currently unclear the overall effect this has on app uptake and use.

Where marketing reaches prospective users, it is expected that a portion will seek out the app. States should work to ensure the conversion from marketing views to app store as simple as possible to minimize the drop off from potential users. Direct links should be used wherever possible; suggesting keywords that have been verified to effectively bring the app can help users make the transition easily.

Leveraging partnerships and fostering positive word-of-mouth are important avenues to ensure a broad reach and positive initial sentiment. It is likely the lessons from consumer marketing in the age of social media - i.e., that recommendations from friends are the most trusted form of advertising, particularly with millennials (McCaskill, 2015; Mulqueen, 2018) - hold true here. States such as Arizona have also indicated positive app uptake by forming public-private partnerships and leveraging commercial marketing reach and brand trust. Engaging trusted community leaders in vulnerable communities is also expected to help app adoption. States should also consider engaging with primary care physicians and labs across their state to coordinate messaging and foster positive word of mouth from people prospective app users are likely to trust for medical advice and medical information.

**Evaluating effects of marketing and outreach.**

- **Focus groups, interviews.** Surveys conducted prior to launch and during app use can be used to identify areas that need to be understood better. Is a certain community or demographic indicating distrust of the apps? Is a certain community or demographic consistently not responding to the surveys? Engaging with community leaders to identify groups of people or individuals to have discussions with can provide the "why" behind the numbers that are being collected, providing the means to target interventions and measure effectiveness.

- **Analytics.** People talk about what is on their mind on social media platforms. This could be a good avenue to gauge initial reactions to state marketing, and whether the campaign is resonating with certain populations.

    o **Metrics.** Number of mentions (key terms or phrases), direct engagement with posts or marketing, positive or negative valence of posts, popularity of misinformation (i.e. shares, likes on posts or comments), geographic location of engagement

- *Optional: Surveys*. Surveys conducted before app deployment can inform initial marketing goals and metrics; surveys conducted after can gauge effectiveness of advertising campaigns.

    o **Metrics.** Number of people indicating awareness of advertising campaigns; cross-section with demographics; state population metrics compared to demographics of survey responses.

**(2) App site visits and downloads.** A percentage of people who find out about the app through any of the outreach methods are expected to visit the app store page for the app. As there are indications that people treat online reviews similarly to recommendations from friends (Mulqueen, 2018), care should be taken with the information presented on the page.

**Evaluating effects of app store page presentation and information.**

- **Analytics.** Though States can't control reviews, they can monitor the reviews for early indications of issues accessing or downloading the app, as well as indications of whether people are finding value in the content. Apple and Google both collect metrics on each app in the app store. States should leverage this information to supplement their own data gathering.
    - **Metrics.** Number of mentions (key terms or phrases), positive or negative valence of posts, popularity of misinformation; Number of visits, number of downloads.
- **Usability tests, Focus groups, A/B testing.** Prior to or after launching content, States can bring variations of app store page layouts and information to prospective users to evaluate attitudes. This can be done in remote online tests, focus groups, or by asking people to compare two different versions to evaluate the relative effect of content on user opinion.
    - **Metrics.** Number of people preferring version A vs B; Average Likert scale rating of trust, understandability, likelihood to download, likelihood to recommend for each version

**(3, 4, 5, 6, 9) In-app communications.** Each user that downloads the app will be presented with a number of screens and options. When developing or evaluating an app, it is important to consider the information presented to users, and understand how the information is perceived. There is a risk/reward calculation that every technology user makes whether they are aware of it or not; if the information from the app is seen as not worth the effort, or worse, is more intrusive than the app is worth, users will discontinue use (see **In-App Trustworthiness Principles**).

**Evaluating effects of in-app communications.**

- **Analytics.** Some apps may collect aggregated usage data that can be leveraged for insight. For instance, Ireland and PA's apps both allow users to opt in to usage data collection, which measures when people open the app and whether they have deleted it.
    - **Metrics.** Number of times opening app, visiting pages; Number of deletions; Number opting in to notifications; Duration of use
- **Survey (in-app).** States that develop an app, or deploy a commercial app have the opportunity to allow users to opt in and provide additional data on their use. PA and Ireland provided the means for users to provide optional in-app demographics, with PA additionally adding an optional in-app survey (both subject to sampling bias, but only source of information).
    - **Metrics.** Number of responses from different demographic areas; Average Likert scale responses to questions on trust, understandability, experience (subject to sampling bias)
- **Usability tests, Focus groups, A/B testing.** Prior to or after launching content, States can bring variations of app workflows and presentation of information to prospective users to evaluate attitudes. This can be done in remote online tests, focus groups, or by asking people to compare two different versions to evaluate the relative effect of content on user opinion.
    - **Metrics.** Number of people preferring version of information; Average Likert scale rating of trust, understandability, experience

**(7) Public health communication – in person.** App experience should be considered together with personal health interactions. Though people that are already feeling sick and getting a test may not be the prime demographic for the app, interactions with personal or public health providers can offer an opportunity for positive communications about the app, as well as opportunities to measure progress.

- **Interviews, focus groups.** Interviews with health providers as well as members of demographics of interest can yield insight into biases, viewpoints, and experiences that may be affecting app perception and use behind the scenes. These communications may also yield ideas for better coordination and messaging throughout States' health infrastructure.
- **Analytics.** States are likely already collecting metrics about COVID from health providers around the States. These metrics should be leveraged and evaluated with app metrics to gain a fuller picture of COVID response.
    - **Metrics**. Number of positive tests referred from different types of health providers (e.g., primary care physicians, urgent care).

**(8, 9) Public health communication – contact tracing call.** People who have a positive COVID test who are also not sick enough to be admitted to a hospital for medical care likely have their first communication about their result through a phone call from a public health official. Delays between a positive test and the phone call can complicate efforts, as can the relative low response rate to phone calls. As a large portion of the apps require users who test positive to upload a key in order to notify their close contacts of a possible exposure, this is a key step in the success of AEN. Care should be taken to ensure that outgoing numbers appear as official public health communications on phones, and that contact tracers have the instructions they need to walk app users through successful loading of keys where required.

- **Interviews.** Interviews with contact tracers and others throughout the process may yield additional insight into their experience, possibly uncovering avenues to improve.
- **Analytics.** States are likely already collecting metrics about COVID from contact tracers. These metrics should be leveraged and evaluated with app metrics to gain a fuller picture of COVID response. Some States have identified potential process issues resulting in keys not being uploaded, resulting in additions made to contact tracing scripts.
    - **Metrics.** How many identified contacts answer phone calls from health providers, how many contacts they in turn provide; how many contacts have the app; how many keys are uploaded; how many phones are contacted from uploaded keys; comparison with positive state test results

## 5   DISCUSSION AND RECOMMENDATIONS

As indicated in a recent MIT Technology Review article, commonly cited cases of successful app roll-out are Germany and Ireland, who have seen download rates of 20% and 37% of their population respectively (Jee, 2020). Though PA and other States have not reached the numbers of Germany and Ireland, there is still work to be done to understand barriers to use and to get the app into more hands. Raw app download numbers are not the full picture; the ideal is to break the chain of spread. Some States, including PA, have identified potential issues in communication or process that could be affecting experience and use. Identifying these issues and mitigating them early is expected to improve app uptake and use. States who have rolled out AEN apps consider that they are preparing all the tools at their disposal to mitigate a second wave of infections. As indicated by the technical director of NearForm, developers of Ireland's and PA's app, "if you break even a few transmission chains as a result of the app, for me that's a success" (Jee, 2020).

**Set measurable goals.** When evaluating AEN roll-out, as with any system where performance is relevant, to go from measurements to actions there must be some set of goals for each metric to reach. By having thresholds to aspire to for each metric, improvement efforts can be targeted to address those goals. Without goals, there is a greater risk of making changes aimlessly; without measurement methods, there is no way to systematically decide which changes to make. In testing different approaches to increasing app enrollment, it is important to carefully consider how success will be measured.

**States need to plan for iterative assessment and improvement.** A central struggle of infectious disease containment is population churn; people who are unexposed that become exposed, people who are pre-infectious that become infectious. The constantly changing set of pre-infectious and infectious people challenges the capacity of public health officials working to contain potential spreaders, and maintaining this capacity is resource intensive. At the same time, this population churn presents the opportunity that a state does not have only a single chance to implement an effective mechanism for controlling infectious disease. The rise and fall of cases and hospitalization rates mean the population may be more or less interested in getting engaged at any point in time. At any time, a state can make progress by measuring the performance of their efforts to identify bottlenecks and make changes to address any factors that limit the ability to control the spread of the virus. In fact, it is crucial for any serious containment measures to evaluate and adjust their efforts to maximize effectiveness. States who are considering AEN app deployments or who have deployed can learn from existing deployments and ongoing assessments.

**Focus on controllable communication points.** Even with the existing body of research, there is much that is unknown about why specific individuals or populations choose to download an AEN app (or not). There are any number of assessments that States can undertake in the drive to improve app performance. In order to make assessments manageable, it is important to target efforts towards areas that are the most likely to have impact. States should look to clarify and reinforce messaging at communication points with app users that are under state control.

**User experience and trust are crucial factors.** Ultimately, the effectiveness of AEN schemes and public health efforts broadly is directly tied to user engagement. For that reason, the user experience needs to be considered at the forefront of the viral containment and app improvement efforts. The methods we summarize in **Human-Focused Assessment Methodologies** may indicate that changes to the app need to be made. Such changes may be technical, but they should also incorporate how any technical details are communicated to any current or potential users through the multitude of communication channels that inform the user (see **Focus Efforts on Controllable Communication Points).** Users need to trust the app and the organizations managing the apps, which carries through in-app communication (see **In-App Trustworthiness Principles**) as well as perceived trustworthiness of communications about the app (Mulqueen, 2018). States should be honest with users about performance expectations, and forthright about any missteps. States should also honor statements about the non-permanent nature of the apps and the protection of user data, and develop a clear exit plan that considers both.

**Human research methods should be leveraged to assess and improve performance.** Human research methods provide States with many tools that should be used to evaluate and improve app adoption and efficacy. We presented a summary of such methods (see **Human-Focused Assessment Methodologies**). As States work to improve use of their deployed systems, a robust selection of methods with different strengths should be used to evaluate adoption and use of the app, as well as user experience and feedback regarding community members' attitudes and behaviors towards such apps. No one method presented offers a complete means of evaluation, so several methods should be taken in conjunction to inform changes to the state's efforts.

# 6 CONCLUSIONS AND NEXT STEPS

There are a number of outstanding questions that have arisen from the PA case study and review of existing deployments and assessments.

A central question is the value proposition of such apps and how they are perceived by the public; does the perceived value of the app outweigh the perceived data sharing risks? A challenge is that the current value of using any given app is largely a function of the percentage of others within a given community who have enrolled; what that percentage will be is not known in advance. It will be incumbent on States to lean heavily on private and public organizations within the State to encourage enrollment in the app among populations that are or are likely to be spread vectors. A second consideration is examining whether introducing features focused on exposure prevention would increase perceived value (e.g., immediate notification of too close for too long contacts, updates to county spread numbers), and through that, expand app use among the population.

Next steps include continued evaluation of user experience with existing apps as discussed above such as identifying populations that are not represented in the apps and deploying outreach, detailing general public understanding of the technology and its effect on behavior, understanding the impact of the apps in identifying positive cases, evaluating what identified cases do and how they behave (i.e. is identifying enough to change behavior), where are we seeing States max out (i.e. what percent population), and identifying outlier States (positive and negative) to study and evaluate. The authors plan to address each point through targeted quantitative surveys and qualitative behavioral studies with individuals. In addition a robust program of A/B testing of different approaches to encourage downloads, as is being undertaken by both the UPenn and MIT/MIT LL teams, is likely to be necessary to both increase app download rates and build evidence on what works in different settings.